\documentclass[11pt]{article}

\usepackage{graphicx,amssymb,amstext,amsmath,amsthm,verbatim,latexsym}  

\usepackage{pstricks,pst-node,pst-tree}

\usepackage{fancyhdr}

\newcommand{\bE}{\mathbb{E}}

\newtheorem{lemma}{Lemma}[section]

\newtheorem{theorem}[lemma]{Theorem}

\newtheorem{claim}[lemma]{Claim}

\newtheorem{corollary}[lemma]{Corollary}

\theoremstyle{definition}

\theoremstyle{definition}

\theoremstyle{remark}

\newtheorem*{remark}{Remark}

\title{Multiparty Communication Complexity of Disjointness}

\author{Arkadev Chattopadhyay and Anil Ada\thanks{authors are supported by research grants of Prof. D. Th{\'e}rien. The first author thanks M. David and T. Pitassi for several discussions.} \\\\
{School of Computer Science} \\
{McGill University, Montreal, Canada} \\
{\small {\tt achatt3,aada\@@cs.mcgill.ca}}
}

\date{}

\begin{document}

\maketitle

\begin{abstract}

We obtain a lower bound of $\Omega\bigg(\frac{n^{\frac{1}{k+1}}}{2^{2^k} (k-1)2^{k-1}}\bigg)$ on the $k$-party randomized communication complexity of the Disjointness function in the `Number on the Forehead' model of multiparty communication. In particular, this yields a bound of $n^{\Omega(1)}$ when $k$ is a constant. The previous best lower bound for three players until recently was $\Omega(\log n)$.

Our bound separates the communication complexity classes $NP^{CC}_k$ and $BPP^{CC}_k$ for $k=o(\log \log n)$. Furthermore, by the results of Beame, Pitassi and Segerlind \cite{BPS07},  our bound implies proof size lower bounds for tree-like, degree $k-1$ threshold systems and superpolynomial size lower bounds for Lov\'{a}sz-Schrijver proofs.

Sherstov \cite{She07b} recently developed a novel technique to obtain lower bounds on two-party communication using the approximate polynomial degree of boolean functions. We obtain our results by extending his technique to the multi-party setting using ideas from Chattopadhyay \cite{Cha07}.   


A similar bound for Disjointness has been recently and independently obtained by Lee and Shraibman.

\end{abstract}

\section{Introduction}

Chandra, Furst and Lipton \cite{CFL83} introduced the `Number on the Forehead' model of multiparty communication as an extension of Yao's \cite{Yao79} two party communication model. This model, besides being interesting in its own right, has found numerous connections with circuit complexity, proof complexity, branching programs, pseudo-random generators and other areas of theoretical computer science. 

Both proving upper and lower bounds for this model remain a very challenging task as it is known that the overlap of information accessible to players provides significant power to it. In fact, proving a super-polylogarithmic lower bound on the communication needed by poly-logarithmic number of players for computing a function $f$ in the restricted setting of simultaneous deterministic communication, is enough to show that $f$ is not in $\text{ACC}^0$, a class for which no strong bounds are known. Although several efforts \cite{BNS92,ChungT,Raz,FG} have been made, this goal currently remains out of reach as no superlogarithmic lower bounds exist for even $\log n$ players.

More modestly, one would like to be able to determine the communication complexity of simple functions for at least constant number of players. However, despite intensive research (see for example \cite{BPSW,C07,VW07,Tesson03}) the best known lower bounds on the communication complexity of simple functions like Disjointness and Pointer Jumping was $\Omega(\log n)$ even for three players. The root cause of this problem is that there was essentially only one method that was the backbone of almost all strong lower bounds. This method is known as the discrepancy method and was introduced in the seminal work of Babai, Nisan and Szegedy \cite{BNS92}. It is however known that for functions like Disjointness this method at best yields $\Omega(\log n)$ lower bounds.

Razborov \cite{Razborov03} introduced the multi-dimensional discrepancy method to establish a tight relationship between the quantum communication complexity of functions induced by a symmetric base function and the approximation degree of the base function. Recently, Sherstov \cite{She07b} develops an elegant technique that is simpler and generalizes the results of Razborov by obviating the need for the base function to be symmetric. More importantly for us, the technique in \cite{She07b} shows that the classical discrepancy method can be modified in a natural way that allows one to obtain strong bounds on two-party quantum communication with bounded error even for functions like Disjointness that have large discrepancy. In this work, we suitably modify this technique to extend it to the multi-party setting. In order to achieve this, we use tools developed in Chattopadhyay \cite{Cha07}, extending the earlier work of Sherstov \cite{She07a}, for estimating discrepancy under certain non-uniform distributions. 


Our result has interesting consequences for communication complexity classes and proof complexity. It provides the first example of an explicit function that has small non-deterministic communication complexity, but exponentially high randomized complexity. In the language of complexity classes, this separates $\text{BPP}^{CC}_k$ and $\text{NP}^{CC}_k$ for $k=o(\log \log n)$. In fact, the separation is exponential when $k$ is any constant. Although such a separation was already known from the work of \cite{BDPW07}, before our work no explicit function was known to separate these classes. By the work of Beame, Pittasi and Szegerlind \cite{BPS07}, our lower bounds on the $k$-party complexity of Disjointness implies strong lower bounds on the proof size for a family of proof systems known as tree-like, degree $k-1$ threshold systems. Proving lower bounds for these systems was a major open problem in propositional proof complexity.

\subsection{Our Main Result}

Let $y^1,\ldots,y^{k-1}$ be $k-1$ $n$-bit binary strings. Define the $k-1 \times n$ boolean matrix $A$ obtained by placing $y^i$ in the $i$th row of $A$. For $x \in \{0,1\}^n$, let $x \Leftarrow y^1,\ldots,y^{k-1}$ be the $n$-bit string $x_{i_1}x_{i_2}\ldots x_{i_t}0^{n-t}$, where $i_1,\ldots, i_t$ are the indices of the all-one columns of $A$.

Let $g:\{0,1\}^n \to \{-1,1\}$ be a \emph{base} function. We define $G^g_k : (\{0,1\}^n)^k \to \{-1,1\}$ by $G^g_k (x, y^1,\ldots,y^{k-1}) := g(x \Leftarrow y^1,\ldots,y^{k-1}$). Observe that $G^{\textrm{PARITY}}_k$ is the Generalized Inner Product function and $G^{\textrm{NOR}}_k$ is the Disjointness function. Our main result shows how to use the high approximation degree of a base function to generate a function with high randomized communication complexity. 

Let $R^{\epsilon}_k(f)$ denote the randomized $k$-party communication complexity of $f$ with advantage $\epsilon$. Then,

\begin{theorem}
Let $f:\{0,1\}^m \to \{-1,1\}$ have $\delta$-approximate degree $d$. Let $n \geq \big ( \frac{2^{2^k} (k-1) e}{d} \big )^{k-1} m^k$, and $f':\{0,1\}^n \to \{-1,1\}$ be such that $f(z) = f'(z0^{n-m})$. Then
\[ R^{\epsilon}_k (G^{f'}_k) \geq \frac{d}{2^{k-1}} + \log(\delta + 2\epsilon - 1). \]
\end{theorem}
\noindent
As a corollary we show that 
\[
 R^\epsilon_k(\text{DISJ}_k) = \Omega\bigg(\frac{n^{\frac{1}{k+1}}}{2^{2^k} (k-1)2^{k-1}}\bigg)
\]
for every constant $\epsilon > 0$. In brief, this follows from the following facts. Let $\textrm{NOR}_n$ denote the NOR function for inputs of length $n$. Then $f' = \textrm{NOR}_n$ and $f = \textrm{NOR}_m$ satisfy $f(z) = f'(z0^{n-m})$ and by a result of Paturi \cite{Pat92}, we know that the $1/3$-approximate degree of $\textrm{NOR}_m$ is $\Theta(\sqrt{m})$.


A similar bound for the Disjointness function has been recently and independently obtained by Lee and Shraibman \cite{LS08}.

\subsection{Proof Overview}

\begin{figure}
\begin{center}
\includegraphics[scale=1]{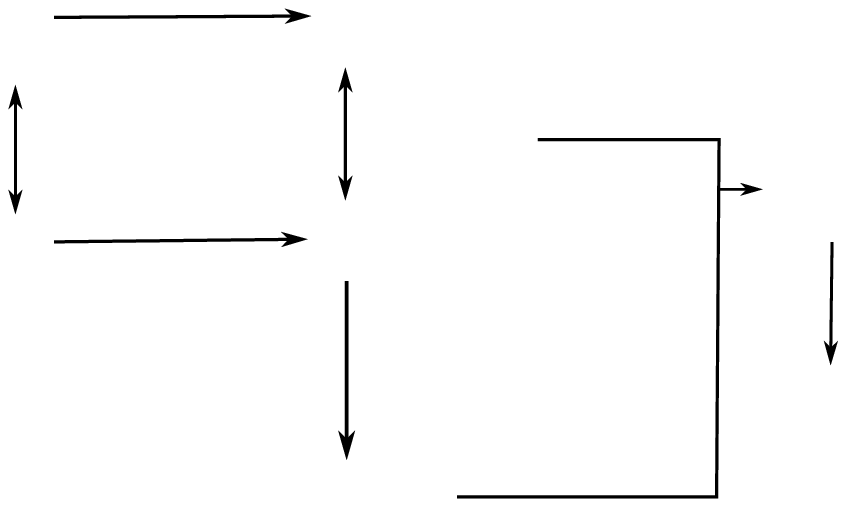}
\rput{0}(-9.8,4.4){Approximation}
\rput{0}(-9.8,3.9){Orthogonality}
\rput{0}(-8.5,5.5){$f$}
\rput{0}(-8.5,5.1){high approx-deg}
\rput{0}(-8.5,3.2){$g$}
\rput{0}(-8.5,2.8){$\bE_{\mu} g(x) p(x) = 0$ for low deg($p$)}
\rput{0}(-5,5.5){$F^f_k$}
\rput{0}(-5,3.2){$F^g_k$}
\rput{0}(-7.5,4.2){high corr.}
\rput{0}(-4.2,4.2){high corr.}
\rput{0}(-5.3,0.5){disc $F^g_k$ is low}
\rput{0}(-6.35,2.1){Orthogonality-}
\rput{0}(-6.35,1.6){Discrepancy}
\rput{0}(-2.5,3.2){Generalized}
\rput{0}(-2.5,2.7){Discrepancy}
\rput{0}(-2.5,2.3){Method}
\rput{0}(0.4,3.7){$R^{\epsilon}_k(F^f_k)$ is high}
\rput{0}(0.4,1.6){$R^{\epsilon}_k(G^f_k)$ is high}
\end{center}
\caption{Proof outline}
\label{outline}
\end{figure}

Sherstov \cite{She07b} devised a novel strategy to make a passage from approximation degree of boolean functions to lower bounds on two-party communication complexity. We adapt this strategy for our purpose. This adaptation is outlined in Figure \ref{outline}.

We use three main ingredients, the first of which is the Generalized Discrepancy Method. The classical discrepancy method states that if a function has low discrepancy, then it has high randomized communication complexity. In the generalized discrepancy method this idea is extended as follows: If a function $g$ correlates well with $f$ and has low discrepancy, then $f$ has high randomized communication complexity.

The second ingredient is the ``Approximation/Orthogonality Principle" of Sherstov \cite{She07b}. It states that given a function $f$ with high approximation degree, we can find a function $g$ that correlates well with $f$, and a distribution $\mu$ such that $g$ is orthogonal to every low degree polynomial under $\mu$.

The third ingredient, called the Orthogonality-Discrepancy Lemma, is derived from the work of Chattopadhyay \cite{Cha07}. This takes a function that is orthogonal with low degree polynomials and constructs a new masked function that has low discrepancy. 

We can then summarize the strategy as follows. We start with a function $f:\{0,1\}^n \to \{-1,1\}$ with high approximation degree. By the Approximation/Orthogonality Principle, we obtain $g$ that highly correlates with $f$ and is orthogonal with low degree polynomials. From $f$ and $g$ we construct new masked functions $F^f_k$ and $F^g_k$, similar to the construction of $G^f_k$. Since $g$ is orthogonal to low degree polynomials, by the Orthogonality-Discrepancy Lemma we deduce that $F^g_k$ has low discrepancy under an appropriate distribution. Under this distribution $F^g_k$ and $F^f_k$ are highly correlated and therefore applying the Generalized Discrepancy Method, we conclude that $F^f_k$ has high randomized communication complexity. This implies, by the construction of $F^f_k$, that the randomized communication complexity of $G^f_k$ is high.

\section{Preliminaries}

\subsection{Multiparty Communication Model}

In the multiparty communication model introduced by \cite{CFL83}, $k$ players $P_1, \ldots , P_k$ wish to collaborate to compute a function $f : \{0,1\}^{n} \rightarrow \{-1,1\}$. The $n$ input bits are partitioned into $k$ sets $X_1, \ldots, X_k \subseteq[n]$ and each participant $P_i$ knows the values of all the input bits {\em except} the ones of $X_i$. This game is often referred to as the ``Number/Input on the forehead'' model since it is convenient to picture that player $i$ has the bits of $X_i$ written on its forehead, available to everyone but itself. Players exchange bits, according to an agreed upon protocol, by writing them on a public blackboard. The protocol specifies whose turn it is to speak, and what the player broadcasts as a function of the communication history and the input the player has access to. The protocol's output is a function of what is on the blackboard after the protocol's termination. We denote by $D_k(f)$ the deterministic $k$-party communication complexity of $f$, i.e.\ the number of bits exchanged in the \emph{best} deterministic protocol for $f$ on the worst case input.

By allowing the players to access a public random string and the protocol to err, one defines the randomized communication complexity of a function. We say that a protocol computes $f$ with $\epsilon$ advantage if the probability that $\mathcal{P}$ and $f$ agree is at least $1/2 + \epsilon$ for all inputs. We denote by $R_k^\epsilon (f)$ the cost of the best protocol that computes $f$ with advantage $\epsilon$. One further introduces non-determinism in protocols by allowing `God' to help the players by furnishing a proof string. As is usual with non-determinism in other models, a correct non-deterministic protocol $\mathcal{P}$ for $f$ has the following property: on every input $x$ at which $f(x)=-1$, $\mathcal{P}(x,y)=-1$ for some proof string $y$ and whenever $f(x)=1$, $\mathcal{P}(x,y)=1$ for all proof strings $y$. The length of the proof string $y$ is now included in the cost of $\mathcal{P}$ on an input and $N_k(f)$ denotes the cost of the best non-deterministic protocol for $f$ on the worst input.

Communication complexity classes were introduced for two players in \cite{BFS86} in which ``efficient" protocol was defined to have cost no more than $polylog(n)$. This idea naturally extends to the multiparty model giving rise to the following classes: $\text{P}^{CC}_k\,:=\,\{f\,|\,D_k(f)\,=\, \text{polylog}(n)\}$, $\text{BPP}^{CC}_k\,:=\,\{f\,|\,R_k^{1/3}(f)\,=\, \text{polylog}(n)\}$ and $\text{NP}^{CC}_k\,:=\,\{f\,|\,N_k(f)\,=\, \text{polylog}(n)\}$. Determining the relationship among these classes is an interesting research theme within the broader area of understanding the relative power of determinism, non-determinism and randomness in computation. While Beame et.al. \cite{BDPW07} show that $\text{BPP}^{CC}_k \ne \text{NP}^{CC}_k$, no explicit function was known that separated these classes.

\subsection{Cylinder Intersections and Discrepancy}

The key combinatorial object that arises in the study of multiparty communication is a \emph{cylinder-intersection}. A $k$-cylinder in the $i$th dimension is a subset $S$ of $Y_1 \times \cdots \times Y_k$ with the property that membership in $S$ is independent of the $i$th coordinate. A set $S$ is called a cylinder-intersection if $S=\cap_{i=1}^k S_i$, where $S_i$ is a cylinder in the $i$th dimension. One can represent a $k$-cylinder in the $i$th dimension by its characteristic function $\phi^i : (\{0,1\}^n)^k \to \{0,1\}$. Here $\phi^i(y_1,...,y_k)$ does not depend on $y_i$. A cylinder intersection is represented as the product 
\[ \phi(y_1,...,y_k) = \phi^1(y_1,...,y_k) ... \phi^k(y_1,...,y_k). \]  

It is well known that a protocol that computes $f$ with cost $c$ partitions the input space of $f$ into at most $2^c$ monochromatic cylinder intersections.

An important measure, defined for a function $f: Y_1 \times ... \times Y_k \to \{-1,1\}$, is its \emph{discrepancy}. With respect to any probability distribution $\mu$ over $Y_1\times\cdots\times Y_k$ and cylinder intersection $\phi$, define 
\begin{eqnarray*}   \label{eq:discrepancydef}
\text{disc}_{k,\mu}^{\phi}(f) & = & \bigg|\Pr_{\mu}\big[f(y_1,\ldots,y_k)=1 \wedge \phi(y_1,\ldots,y_k)=1\big] \\
& & - \Pr_{\mu}\big[f(y_1,\ldots,y_k)=-1 \wedge \phi(y_1,\ldots,y_k)=1\big]\bigg|.
\end{eqnarray*}
Since $f$ is -1/1 valued, it is not hard to verify that equivalently:
\begin{align}   \label{eq:discrepancydef1}
\text{disc}_{k,\mu}^{\phi}(f) = \bigg|\mathbb{E}_{y_1,\ldots,y_k \sim \mu}f(y_1,\ldots,y_k)\phi(y_1,\ldots,y_k)\bigg|.
\end{align}

The discrepancy of $f$ w.r.t. $\mu$, denoted by $\text{disc}_{k,\mu}(f)$ is $\text{max}_{\phi} \text{disc}_{k,\mu}^{\phi}(f)$. For removing notational clutter, we often drop $\mu$ from the subscript when the distribution is clear from the context. We now state the discrepancy method which connects the discrepancy and the randomized communication complexity of a function.

\begin{theorem}[see \cite{BNS92,KN97}]     \label{theorem:discrepancytocomplexity}
Let $0 < \epsilon \leq 1/2$ be any real and $k \ge 2$ be any integer. For every function $f:Y_1 \times ... \times Y_k \to \{1,-1\}$ and distribution $\mu$ on inputs from $Y_1 \times \cdots \times Y_k$,
\begin{align}   \label{eq:randomized-discrepancy}
R_k^{\epsilon}(f) \ge \log\bigg(\frac{2\epsilon}{\text{disc}_{k,\mu}(f)}\bigg).
\end{align} 
\end{theorem}

\subsection{Fourier Expansion}

We consider the vector space of functions from $\{0,1\}^n \rightarrow \mathbb{R}$. Equip this space with the standard inner product $\langle f,g \rangle$
\begin{align}   \label{eq:innerproduct}
\langle f,g \rangle = \mathbb{E}_{x \sim \mathcal{U}} f(x)g(x)
\end{align}

For each $S\subseteq [n]$, define $\chi_S(x) = (-1)^{\sum_{i\in S} x_i}$. Then it is easy to verify that the set of functions $\{\chi_S | S \subseteq [n] \}$ forms an orthonormal basis for this inner product space, and so every $f$ can be expanded in terms of its \emph{Fourier coefficients} 
\begin{align}   \label{eq:fourierexpansion}
f(x) = \sum_{S \subseteq [n]} \hat{f}(S) \chi_S(x)
\end{align}
where $\hat{f}(S)$ is defined as $\langle f,\chi_S \rangle$. This expansion is unique and the \emph{exact degree} of $f$ is defined to be the largest $d$ such that there exists $S\subseteq [n]$ with $|S|=d$ and $\hat{f}(S) \ne 0$.

\subsection{Approximation Degree}

A natural question is the following. How large degree is needed if we want to simply approximate $f$ well? Define the $\epsilon$-\emph{approximate degree of $f$}, denoted by $\text{deg}_{\epsilon}(f)$ to be the smallest integer $d$ for which there exists a function $\phi$ of exact degree $d$ such that 
\begin{align} 
\text{max}_{x \in \{0,1\}^n} \bigg|f(x)-\phi(x)\bigg| \le \epsilon  \nonumber
\end{align}

For any $D:\{0,1,\ldots,n\} \rightarrow \{1,-1\}$, define
\[ \ell_0(D)  \in \{0,1,\ldots,\lfloor n/2 \rfloor \} \]
\[ \ell_1(D)  \in \{0,1,\ldots,\lceil n/2 \rceil \} \]
such that $D$ is constant over the interval $[\ell_0(D),n-\ell_1(D)]$ and $\ell_0(D)$ and $\ell_1(D)$ are the smallest possible values for which this happens.

Paturi's theorem provides bounds on the approximate degree of symmetric functions.

\begin{theorem}[Paturi\cite{Pat92}]    \label{theorem:paturi}
Let $f:\{0,1\}^n \rightarrow \{1,-1\}$ be any symmetric function induced from the predicate $D:\{0,\ldots,n\}\rightarrow\{1,-1\}$. Then,
\begin{align}
\text{deg}_{1/3}(f) = \Theta \big(\sqrt{n(\ell_0(D)+\ell_1(D))}\big)
\end{align}
\end{theorem}

In particular, the 1/3-approximate degree of NOR is $\Theta(\sqrt{n})$.

\section{The Generalized Discrepancy Method}

Babai, Nisan and Szegedy \cite{BNS92} estimated the discrepancy of functions like $\text{GIP}_k$ w.r.t $k$-wise cylinder intersections and the uniform distribution. These estimates resulted in the first strong lower bounds in the k-party model via Theorem~\ref{theorem:discrepancytocomplexity}. Unfortunately, the applicability of Theorem~\ref{theorem:discrepancytocomplexity} is limited to those functions that have small discrepancy. Disjointess is a classical example of a function that does not have small discrepancy.

\begin{lemma}[Folklore]
Under every distribution $\mu$ over the inputs, 
\[ disc_{k,\mu}(DISJ_k) = \Omega(1/n).\]
\end{lemma}
\begin{proof}
Let $X^{+}$ and $X^{-}$ be the set of disjoint and non-disjoint inputs respectively. The first thing to observe is that if $|\mu(X^{+})-\mu(X^{-})|=\Omega(1/n)$, then we are done immediately by considering the discrepancy over the intersection corresponding to the entire set of inputs. Hence, we may assume $|\mu(X^{+})-\mu(X^{-})|=o(1/n)$. Thus, $\mu(X^{-}) \ge 1/2 - o(1/n)$. However, $X^{-}$ can be covered by the following $n$ \emph{monochromatic} cylinder intersections: let $C_i$ be the set of inputs in which the $i$th column is an all-one column. Then $X^{-} = \cup_{i=1}^n C_i$. By averaging, there exists an $i$ such that $\mu(C_i) \ge 1/2n - o(1/n^2)$. Taking the discrepancy of this $C_i$, we are done.
\end{proof}

It is therefore impossible to obtain better than $\Omega(\log n)$ bounds on the communication complexity of Disjointness by a direct application of the discrepancy method. In fact, the above argument shows that Theorem~\ref{theorem:discrepancytocomplexity} fails to give better than polylogarithmic lower bound for every function that is in $\text{NP}^{CC}_k$ or $\text{co-NP}^{CC}_k$.

Sherstov [16, Sec 2.4] 
provides a nice reinterpretation of Razborov's discrepancy method for two party quantum communication complexity by pointing out the following:  in order to prove a lower bound on the communication complexity of a function $f$ in any bounded error model, it is sufficient to find a function $g$ that correlates well with $f$ under some distribution but has large communication complexity. Based on this observation, we modify the discrepancy method to the following:


\begin{lemma}[Generalized Discrepancy Method] \label{lemma:generalized_discrepancy}
Denote $X = Y_1 \times ... \times Y_k$. Let $f:X \rightarrow \{-1,1\}$ and $g:X \rightarrow \{-1,1\}$ be such that under some distribution $\mu$ we have $\text{Corr}_{\mu}(f,g) \geq \delta$. Then 
\begin{align}  \label{eq:gen_disc_0}
R_k^{\epsilon}(f) \geq \log \bigg(\frac{\delta + 2\epsilon - 1}{\text{disc}_{k,\mu}(g)}\bigg)
\end{align} 
\end{lemma}
\begin{proof}
Let $\mathcal{P}$ be a $k$-party randomized protocol that computes $f$ with advantage $\epsilon$ and cost $c$. Then for every distribution $\mu$ over the inputs, we can derive a deterministic $k$-player protocol $\mathcal{P}'$ for $f$ that errs only on at most $1/2 - \epsilon$ fraction of the inputs (w.r.t. $\mu$) and has cost $c$. Take $\mu$ to be a distribution satisfying the correlation inequality. We know $\mathcal{P}'$ partitions the input space into at most $2^c$ monochromatic (w.r.t. $\mathcal{P}'$) cylinder intersections. Let $\mathcal{C}$ denote this set of cylinder intersections. Then,
\begin{eqnarray*}
\delta & \leq & \big| \mathbb{E}_{x \sim \mu} f(x)g(x) \big| \\
         & =    & \big| \sum_x f(x) g(x) \mu(x) \big| \\
	 & \leq & \big| \sum_x \mathcal{P}'(x)g(x)\mu(x)\big| + \big|\sum_x (f(x) - \mathcal{P}'(x)) g(x) \mu(x) \big|
\end{eqnarray*}
Since $\mathcal{P}'$ is a constant over every cylinder intersection $S$ in $\mathcal{C}$, we have
\begin{eqnarray*}
\delta    & \leq & \sum_{S \in \mathcal{C}} \big| \sum_{x \in S} \mathcal{P}'(x) g(x) \mu(x) \big| + \sum_x \big| g(x) \big| \big| f(x) - \mathcal{P}'(x) \big| \mu(x) \\
	       & \leq & \sum_{S \in \mathcal{C}} \big| \sum_{x \in S} g(x) \mu(x) \big| + \sum_x \big| f(x) - \mathcal{P}'(x) \big| \mu(x) \\
	    & \leq & 2^c \text{disc}_{k,\mu}(g) + 2(1/2 - \epsilon).
\end{eqnarray*}
This gives us immediately \eqref{eq:gen_disc_0}.
\end{proof}

Observe that when $f = g$, i.e. $\text{Corr}_\mu (f,g) = 1$, we get the classical discrepancy method (Theorem \ref{theorem:discrepancytocomplexity}).

\section{Generating Functions With Low Discrepancy}

\subsection{Masking Schemes}

We have already defined one masking scheme through the notation $x \Leftarrow y_1,\ldots,y_k$. This allowed us to define $G^g_k$ for a base function $g$. Well-known functions such as $\textrm{GIP}_k$ and $\textrm{DISJ}_k$ are respresentable in this notation by
$G^{\textrm{PARITY}}_k$ and $G^{\textrm{NOR}}_k$ respectively. We now define a second masking scheme which plays a crucial role in lowerbounding the communication complexity of $G^g_k$. This masking scheme is obtained by first slightly simplifying the pattern matrices in \cite{She07b} and then generalizing the simplified matrices to higher dimension for dealing with multiple players.

\begin{figure}
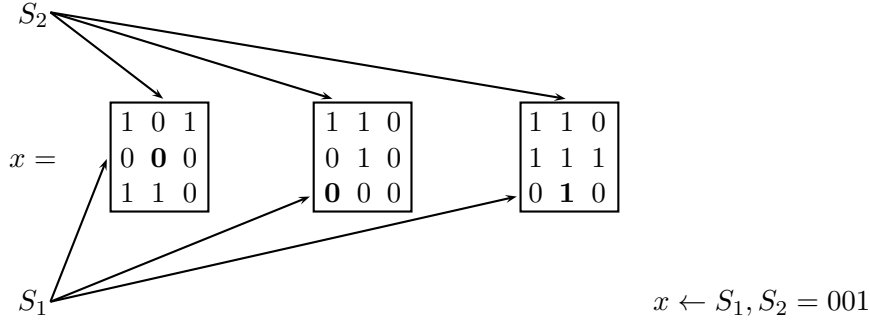

\[
\begin{psmatrix}%
[rowsep=0pt, colsep=6pt]
S_2&&&&&&&&&&&&&&&&&&&&&&&&&&&&&\\
&&&&&&&&&&&&&&&&&&&&&&&&&&&&&\\
&&&&&&&&&&&&&&&&&&&&&&&&&&&&&\\
&&&&1 & 0 & 1 & & & & & & & & 1 & 1 & 0 & & & & & & & & 1 & 1 & 0 &&&\\
x = &&&&0 & \mathbf{0} & 0 & & & & & & & & 0 & 1 & 0 & & & & & & & & 1 & 1 & 1 &&&\\
&&&&1 & 1 & 0 & & & & & & & & \mathbf{0} & 0 & 0 & & & & & & & & 0 & \mathbf{1} & 0 &&&\\
&&&&&&&&&&&&&&&&&&&&&&&&&&&&&\\
&&&&&&&&&&&&&&&&&&&&&&&&&&&&&\\
S_1 &&&&&&&&&&&&&&&&&&&&&&&&&&&&& x \leftarrow S_1,S_2 = 001\\
\end{psmatrix}
\pspolygon(-4.65,2.65)(-3.35,2.65)(-3.35,1.2)(-4.65,1.2)
\pspolygon(-7.4,2.65)(-6.1,2.65)(-6.1,1.2)(-7.4,1.2)
\pspolygon(-10.1,2.65)(-8.8,2.65)(-8.8,1.2)(-10.1,1.2)
\psline{->}(-10.9,0)(-10.15,1.9)
\psline{->}(-10.9,0)(-7.45,1.4)
\psline{->}(-10.9,0)(-4.7,1.4)
\psline{->}(-10.9,3.85)(-9.4,2.7)
\psline{->}(-10.9,3.85)(-7.15,2.7)
\psline{->}(-10.9,3.85)(-4.05,2.7)
\]
\caption{Illustration of the masking scheme $x \leftarrow S_1,S_2$. The parameters are $\ell = 3, m=3, n=27$.}
\label{mask}
\end{figure}

Let $S^1,\ldots S^{k-1} \in [\ell]^m$ for some positive $\ell$ and $m$. Let $x \in \{0,1\}^n$ where $n = \ell^{k-1}m$. Here it is convenient to think of $x$ to be divided into $m$ equal blocks where each block is a $k-1$-dimensional array with each dimension having size $\ell$. Each $S^i$ is a vector of length $m$ with each co-ordinate being an element from $\{1,\ldots,\ell\}$. The $k-1$ vectors $S^1,\ldots,S^{k-1}$ jointly unmask $m$ bits of $x$, denoted by $x \leftarrow S^1,\ldots,S^{k-1}$,  precisely one from each block of $x$ i.e.
\[x[1][S^1[1],S^2[1],...,S^{k-1}[1]],\ldots, x[m][S^1[m],S^2[m],\ldots,S^{k-1}[m]].\]
where $x[i]$ refers to the $i$th block of $x$. See Figure \ref{mask} for an illustration of this masking scheme.

For a given base function $f:\{0,1\}^m \to \{-1,1\}$, we define $F^{f}_k : \{0,1\}^n \times ([\ell]^m)^{k-1} \rightarrow \{-1,1\}$ as $F^{f}_k (x,S^1,\ldots,S^{k-1}) = f(x \leftarrow S^1,\ldots,S^{k-1})$.

\begin{lemma}\label{promise}
If $f: \{0,1\}^m \to \{-1,1\}$ and $f':\{0,1\}^n \to \{-1,1\}$ have the property that $f(z) = f'(z0^{n-m})$ (here $n = \ell^{k-1}m$ as described in the construction of $F^{f}_k$), then
\begin{equation}\label{submatrix}
 R^{\epsilon}_k (F^{f}_k) \leq R^{\epsilon}_k (G^{f'}_k).
\end{equation}
\end{lemma}
\begin{proof}[Proof Sketch]
Observe that there are functions $\Gamma_i : [\ell]^m \to \{0,1\}^n$ such that $F^{f}_k(x, S^1,\ldots,S^{k-1}) = G^{f'}_k(x,\Gamma_1(S^1),\ldots,\Gamma_{k-1}(S^{k-1}))$ for all $x,S^1,\ldots,S^{k-1}$. Therefore the players can privately convert their inputs and apply the protocol for $G^{f'}_k$.
\end{proof}

Note that the proof shows \eqref{submatrix} holds not just for randomized but any model of communication.

\subsection{Orthogonality and Discrepancy}

Now we prove that if the base function $f$ in our masking scheme has a certain nice property, then the masked function $F_k^f$ has small discrepancy. To describe the nice property, let us define the following: for a distribution $\mu$ on the inputs, $f$ is $(\mu,d)$-orthogonal if $\mathbb{E}_{x \sim \mu} f(x)\chi_S(x)=0$, for all $|S|<d$. Then,


\begin{lemma}[Orthogonality-Discrepancy Lemma] \label{deg-disc}
Let $f:\{-1,1\}^m \rightarrow \{-1,1\}$ be any $(\mu,d)$-orthogonal function for some distribution $\mu$ on $\{-1,1\}^m$ and some integer $d > 0$.
Derive the probability distribution $\lambda$ on $\{-1,1\}^{n} \times \big([\ell]^m\big)^{k-1}$ from $\mu$ as follows: $\lambda(x,S^1,\ldots,S^{k-1}) = \frac{\mu(x \leftarrow S^1,\ldots,S^{k-1})}{\ell^{m(k-1)}2^{n-m}}$. Then,
\begin{align}     \label{eq:main}
\bigg(\text{disc}_{k,\lambda}\big(F^f_{k}\big)\bigg)^{2^{k-1}} \le \sum_{j=d}^{(k-1)m} {(k-1)m \choose j}\bigg(\frac{ 2^{2^{k-1}-1}}{\ell - 1}\bigg)^j 
\end{align}
Hence, for $\ell-1 \ge \frac{2^{2^k}(k-1)em}{d}$ and $d>2$, 
\begin{align}    \label{eq:maindisc}
\text{disc}_{k,\lambda}\big(F^f_{k}\big) \le \frac{1}{2^{d/2^{k-1}}}.
\end{align}
\end{lemma}
\begin{remark}
The Lemma above appears very similar to the Multiparty Degree-Discrepancy Lemma in \cite{Cha07} that is an extension of the two party Degree-Discrepancy Theorem of \cite{She07a}. There, the magic property on the base function is high voting degree. It is worth noting that $(\mu,d)$-orthogonality of $f$ is equivalent to voting degree of $f$ being at least $d$. Indeed the proof of the above Lemma is almost identical to the proof of the Degree-Discrepancy Lemma save for the minor details of the difference between our masking scheme and the one used in \cite{Cha07}. 
\end{remark}


\begin{proof}[Proof of Lemma \ref{deg-disc}]
The starting point is to write the expression for discrepancy w.r.t. an arbitrary cylinder intersection $\phi$,
\begin{align}  \label{eq:kcylindricaldisc1}
\text{disc}_k^{\phi}(F^f_k) = \bigg|\sum_{x,S^1,\ldots,S^{k-1}} F^f_k(x,S^1,\ldots,S^{k-1})\phi(x,S^1,\ldots,S^{k-1}) \cdot \lambda(x,S^1,\ldots,S^{k-1})\bigg| 
\end{align}
This changes to the more convenient expected value notation as follows:
\begin{align}  \label{eq:kcylindricaldisc2}
\text{disc}_k^{\phi}(F_k^f) = 2^m\bigg|\bE_{x,S^1,\ldots,S^{k-1}} F^f_k(x,S^1,\ldots,S^{k-1})\times\; \phi(x,S^1,\ldots,S^{k-1})\mu\big( x \leftarrow S^1,\ldots,S^{k-1}  \big) \bigg|
\end{align}
where, $(x,S^1,\ldots,S^{k-1})$ is now uniformly distributed over $\{0,1\}^{\ell^{k-1}m} \times \big([\ell]^m\big)^{k-1}$. Then, we use the trick of repeatedly combining triangle inequality with Cauchy-Schwarz exactly as done in Chattopadhyay\cite{Cha07} (or even before by Raz\cite{Raz}) to obtain the following:
\begin{align}  \label{eq:cylindricaldisck}
(\text{disc}_k^{\phi}(F^f_k))^{2^{k-1}} \le 2^{2^{k-1}m} \bE_{S^1_0,S^1_1,\ldots,S^{k-1}_0,S^{k-1}_1} H_k^f\big(S^1_0,S^1_1,\ldots,S^{k-1}_0,S^{k-1}_1\big)
\end{align}
where,
\begin{align}   \label{eq:cylindricaldisck-1}
& H_k^f\big(S^1_0,S^1_1,\ldots,S^{k-1}_0,S^{k-1}_1\big) & \nonumber \\
& = \bigg|\bE_{x \in \{0,1\}^{\ell^{k-1}m}} \prod_{u\in\{0,1\}^{k-1}} \bigg( F^f_k(x,S^1_{u_1},\ldots,S^{k-1}_{u_{k-1}}) \mu(x \leftarrow S^1_{u_1},\ldots,S^{k-1}_{u_{k-1}}) \bigg) \bigg| &
\end{align}

We look at a fixed $S^i_0,S^i_1$, for $i=1,\ldots,k-1$. Let $r_i=\big|S^i_0 \cap S^i_1\big|$ and $r=\sum_i r_i$ for $1\le i \le 2^{k-1}$. We now make two claims that are analogous to Claim 15 and Claim 16 respectively in \cite{Cha07}.

\begin{claim}  \label{claim:kplayer2}
\begin{align}  \label{eq:kplayer2}
H_k^f\big(S^1_0,S^1_1,\ldots,S^{k-1}_0,S^{k-1}_1\big) \le \frac{2^{(2^{k-1}-1)r}}{2^{2^{k-1}m}}
\end{align}
\end{claim}

\begin{claim}  \label{claim:kplayer1}
Let $r < d$. Then,
\begin{align}  \label{eq:kplayer1}
H_k^f\big(S^1_0,S^1_1,\ldots,S^{k-1}_0,S^{k-1}_1\big) = 0
\end{align}
\end{claim}

We prove these claims in the next section. Claim~\ref{claim:kplayer2} simply follows from the fact that $\mu$ is a probability distribution and $f$ is 1/-1 valued while Claim~\ref{claim:kplayer1} uses the $(\mu,d)$-orthogonality of $f$. We now continue with the proof of the Orthogonality-Discrepancy Lemma assuming these claims. Applying them, we obtain
\begin{align}   \label{eq:kplayer3}
& (\text{disc}_k^{\phi}(F_k^f))^{2^{k-1}} & \nonumber \\ 
& \le \sum_{j=d}^{(k-1)m} 2^{(2^{k-1}-1)j} 
  \sum_{j_1+\cdots+j_{k-1}=j}\Pr\big[r_1 = j_1 \wedge \cdots \wedge r_{k-1} = j_{k-1}\big] &
\end{align}
Substituting the value of the probability, we further obtain:
\begin{align}   \label{eq:kplayer4}
& (\text{disc}_k^{\phi}(F_k^f))^{2^{k-1}} & \nonumber \\
& \le \sum_{j=d}^{(k-1)m} 2^{(2^{k-1}-1)j} 
  \sum_{j_1+\cdots+j_{k-1}=j}{m \choose j_1}\cdots{m \choose j_{k-1}}\frac{(\ell-1)^{m-j_1}\cdots(\ell-1)^{m-j_{k-1}}}{\ell^{(k-1)m}} &
\end{align}

The following simple combinatorial identity is well known: 
\[ \sum_{j_1+\cdots+j_{k-1}=j}{m \choose j_1}\cdots{m \choose j_{k-1}} = {(k-1)m \choose j} \]

Plugging this identity into \eqref{eq:kplayer4} immediately yields \eqref{eq:main} of the Orthogonality-Discrepancy Lemma. Recalling ${(k-1)m \choose j} \leq \big(\frac{e(k-1)m}{j}\big)^j$, and choosing $\ell-1 \geq 2^{2^k}(k-1)em/d$, we get \eqref{eq:maindisc}.
\end{proof}

\subsection{Proofs of Claims}

We identify the set of all assignments to boolean variables in $X=\{x_1,\ldots,x_n\}$ with the $n$-ary boolean cube $\{0,1\}^n$. For any $u \in \{0,1\}^{k-1}$, let $Z_u$ represent the set of $m$ variables indexed jointly by $S^1_{u_1},\ldots,S^{k-1}_{u_{k-1}}$. There is precisely one variable chosen from each block of $X$. Denote by $Z_i[\alpha]$ the unique variable in $Z_i$ that is in the $\alpha$th block of $X$, for each $1 \le \alpha \le m$. Let $Z = \cup_u Z_u$. We abuse notation for the sake of clarity and use $Z_u$ in the context of expected value calculations to also mean a uniformly chosen random assignment to the variables in the set $Z_u$.  

\begin{proof}[Proof of Claim~\ref{claim:kplayer1}]
\begin{align}   \label{eq:claim1-1}
& H_k^f\big(S^1_0,S^1_1,\ldots,S^{k-1}_0,S^{k-1}_1\big) \nonumber \\
& = \bigg|\bE_{Z_{0^{k-1}}}f(Z_{0^{k-1}})\mu(Z_0) \; \bE_{X-Z_{0^{k-1}}} \prod_{\substack{ u\in \{0,1\}^{k-1} \\ u \neq 0}} f(Z_u) \mu(Z_u)  \bigg| 
\end{align}

Observe that for any block $\alpha$ and any $u \neq 0^{k-1}$,  $Z_u[\alpha] = Z_{0^{k-1}}[\alpha]$ iff for each $i$ such that $u_i = 1$, $S^i_0[\alpha] = S^i_1[\alpha]$. Recall that $r_i$ is the number of indices $\alpha$ such that $S^i_0[\alpha] = S^i_1[\alpha]$. Therefore, there are at most $r=\sum_{i=1}^{k-1} r_i$ many indices $\alpha$ such that $Z_u[\alpha] = Z_{0^{k-1}}[\alpha]$ for some $u \neq 0^{k-1}$. This means the inner expectation in \eqref{eq:claim1-1} is a function that depends on at most $r$ variables. Since $f$ is orthogonal under $\mu$ with every polynomial of degree less than $d$ and $r<d$, we get the desired result. 
\end{proof}

\begin{proof}[Proof of Claim~\ref{claim:kplayer2}]
Observe that since $F^f_k$ is 1/-1 valued, we get the following:
\begin{align}  
 H_k^f\big(S^1_0,S^1_1,\ldots,S^{k-1}_0,S^{k-1}_1\big) & \leq \bE_{x} \prod_{u\in\{0,1\}^{k-1}} \mu(x \leftarrow S^1_{u_1},\ldots,S^{k-1}_{u_{k-1}}) \nonumber \\
& = \bE_{X-Z} \; \bE_{Z} \prod_{u \in \{0,1\}^{k-1}} \mu(Z_u) \nonumber \\
& = \bE_{X-Z} \; \frac{1}{2^{|Z|}} \; \sum_{Z \in \{0,1\}^{|Z|}} \; \prod_{u\in\{0,1\}^{k-1}} \mu(Z_u) \label{claim2:1} \\
& \leq  \bE_{X-Z} \; \frac{1}{2^{|Z|}} \; \sum_{\substack{y^1,\ldots,y^{k-1} \\ \in \{0,1\}^m}} \;\; \prod_{i=1}^{k-1} \mu(y^i) \label{claim2:2}
\end{align}
where the last inequality holds because every product in the inner sum of \eqref{claim2:1} appears in the inner sum of \eqref{claim2:2}. Using the fact that $\mu$ is a probability distribution, we get:
\begin{align*}
\text{RHS of \eqref{claim2:2}} & = \bE_{X-Z} \; \frac{1}{2^{|Z|}} \; \prod_{i=1}^{k-1} \sum_{y^i \in \{0,1\}^m} \mu(y^i) \\
& = \bE_{X-Z} \; \frac{1}{2^{|Z|}} \\
& =  \frac{1}{2^{|Z|}}.
\end{align*}

We now find a lower bound on $|Z|$. Let $t_u$ denote the Hamming weight of the string $u$ and $\{j_1, \ldots, j_{t_u}\}$ denote the set of indices in $[k-1]$ at which $u$ has a 1. Define
\begin{align}
Y_u =\big \{ Z_u[\alpha]\;|\; S^{j_s}_1[\alpha] \neq S^{j_s}_0[\alpha];\, 1\le s \le t_i;\, 1 \le \alpha \le m\big\}
\end{align}

The following follow from the above definition.
\begin{itemize}
\item $|Y_{0^{k-1}}|=m $ and $|Y_u| \geq m - \sum_{1\le s \le t_i} r_{j_s} \geq m -r$ for all $u \neq 0^{k-1}$.
\item $Y_u \cap Y_{v} = \emptyset$, for $u \ne v$. This follows from the following argument: wlog assume there is an index $\beta$ where $u$ has a one but $v$ has a zero. Consider any block $\alpha$ such that $Z_u[\alpha]$ is in $Y_u$. It must be true that $S^{\beta}_1[\alpha] \neq S^{\beta}_0[\alpha]$. This means that $Z_u[\alpha] \neq Z_{v}[\alpha]$. Therefore $Z_u[\alpha]$ is not in $Y_{v}$ and we are done.
\item $Y:=\cup_{u \in \{0,1\}^{k-1}} Y_u = Z$. This is because if $Z_u[\alpha]$ is not in $Y_u$ then there are indices $j_1,\ldots,j_s$ where $u$ contains a one and $S_0^{j_i}[\alpha] = S_1^{j_i}[\alpha]$. Let $v$ be the string that contains a zero at positions $j_1,\ldots,j_s$ and at other positions, corresponds to $u$. Then by definition, $Z_u[\alpha] = Z_v[\alpha] \in Y_v$.
\end{itemize}

Thus, $|Z| = |Y| = \sum_{u} |Y_u| \geq m + \sum_{u \neq 0}(m-r) = 2^{k-1}m-(2^{k-1}-1)r$ and the result follows.
\end{proof}

\section{The Main Result}

Before proving the main result, we borrow from Sherstov \cite{She07b} a beautiful duality between approximability and orthogonality. The intuition is that if a function is at a large distance from the linear space spanned by the characters of degree less than $d$, then its projection on the dual space spanned by characters of degree at least $d$ is large. More precisely, 

\begin{lemma}\label{lemma:approximation}
Let $f:\{-1,1\}^m \rightarrow \mathbb{R}$ be given with $\text{deg}_{\delta}(f) = d \geq 1$. Then there exists $g: \{-1,1\}^m \to \{-1,1\}$ and a distribution $\mu$ on $\{-1,1\}^m$ such that $g$ is $(\mu,d)$-orthogonal and $\text{Corr}_{\mu} (f , g) > \delta$.
\end{lemma}

We do not prove this Lemma but the interested reader can read its short proof in \cite{She07b} which is based on an application of linear programming duality. 

\begin{theorem}[Main Theorem] \label{theorem:main}
Let $f:\{0,1\}^m \to \{-1,1\}$ have $\delta$-approximate degree $d$. Let $n \geq \big ( \frac{2^{2^k} (k-1) e}{d} \big )^{k-1} m^k$, and $f':\{0,1\}^n \to \{-1,1\}$ be such that $f(z) = f'(z0^{n-m})$. Then
\begin{align}\label{main:statement}
R^{\epsilon}_k (G^{f'}_k) \geq \frac{d}{2^{k-1}} + \log(\delta + 2\epsilon - 1). 
\end{align}
\end{theorem}
\begin{proof}
Applying Lemma~\ref{lemma:approximation} we obtain a function $g$ and a distribution $\mu$ such that $\text{Corr}_{\mu}(f, g) > \delta$ and $\mathbb{E}_{x \sim \mu} g(x) \chi_S(x) = 0 \text{  for }|S|<d$. These $g$ and $\mu$ satisfy the conditions of Lemma~\ref{deg-disc}, therefore we have
\begin{align} \label{main:disc} 
\text{disc}_{k,\lambda}\big(F^g_{k}\big) \le \frac{1}{2^{d/2^{k-1}}} 
\end{align}
where $\lambda$ is obtained from $\mu$ as stated in Lemma~\ref{deg-disc} and $\ell \ge 2^{2^k}(k-1)em/d$. Since $n = \ell^{k-1}m$, \eqref{main:disc} holds for $n \geq \big (\frac{2^{2^k} (k-1) e}{d} \big)^{k-1} m^k$.

It can be easily verified that $\text{Corr}_{\lambda}(F^{f}_k, F^g_k) = \text{Corr}_{\mu}(f, g) > \delta$. Thus, by plugging the value of $\text{disc}_{k,\lambda}\big(F^g_{k}\big)$ in (\ref{eq:gen_disc_0}) of the generalized discrepancy method we get 
\[ R^{\epsilon}_k(F^{f}_k) \geq \frac{d}{2^{k-1}} + \log(\delta + 2\epsilon - 1). \]
The desired result is obtained by applying Lemma~\ref{promise}.
\end{proof}

\subsection{Disjointness Separates $\text{BPP}^{CC}_k$ and $\text{NP}^{CC}_k$}

As a corollary to our main theorem, we obtain the following lower bound for the Disjointness function.

\begin{corollary}
\[ R_k^{\epsilon} (\text{DISJ}_k) = \Omega\bigg(\frac{n^{\frac{1}{k+1}}}{2^{2^k} (k-1)2^{k-1}}\bigg) \] for any constant $\epsilon > 0$.
\end{corollary}
\begin{proof}
Let $f = \text{NOR}_m$ and $f' = \text{NOR}_n$. We know $\text{deg}_{1/3}(\text{NOR}_m) = \Theta(\sqrt{m})$ by Theorem~\ref{theorem:paturi}. Setting $n = \big (\frac{2^{2^k} (k-1) e}{\text{deg}_{1/3}(\text{NOR}_m)} \big)^{k-1} m^k$, and writing \eqref{main:statement} in terms of $n$ gives the result for any constant $\epsilon > 1/6$. The bound can be made to work for every constant $\epsilon$ by a standard boosting argument.
\end{proof}

Observe that we get the same bound for the function $G^\text{OR}_k$. It is not difficult to see that there is a $O(\log n)$ bit non-deterministic protocol for $G^\text{OR}_k$ and therefore this function separates the communication complexity classes $\text{BPP}^{CC}_k$ and $\text{NP}^{CC}_k$ for all $k=o(\log \log n)$.

\subsection{Other Symmetric Functions}

Theorem~\ref{theorem:main} does not immediately provide strong bounds on the communication complexity of $G^f_k$ for every symmetric $f$. For instance, if $f$ is the MAJORITY function then one has to work a little more to derive strong lower bounds.

In this section, using the main result and Paturi's Theorem (Theorem~\ref{theorem:paturi}), we obtain a lower bound on the communication complexity of $G^f_k$ for each symmetric $f$. Let $f: \{0,1\}^n \to \{1,-1\}$ be the symmetric function induced from a predicate $D:\{0,1,\ldots,n\} \to \{1,-1\}$. We denote by $G^D_k$ the function $G^f_k$. For $t \in \{0,1,\ldots,n-1\}$, define $D_t:\{0,1,\ldots,n-t\} \to \{1,-1\}$ by $D_t(i) = D(i+t)$. Observe that the communication complexity of $G^{D}_k$ is at least the communication complexity of $G^{D_t}_k$.

\begin{corollary}
Let $D:\{0,1,\ldots,n\}$ be any predicate with $\text{deg}_{1/3}(D) = d$. Let $\ell_0 = \ell_0(D)$ and $\ell_1 = \ell_1(D)$. Define $T:\mathbb{N} \to \mathbb{N}$ by 
\[ T(n) = \bigg( \frac{n}{ (2^{2^k} (k-1) e /d)^{k-1} } \bigg)^{\frac{1}{k}} \]
Then for any constant $\epsilon > 0$,
\[ 
R^{\epsilon}_k (G^D_k) = \Omega\bigg(\Psi(\ell_0) + \frac{T(\ell_1)}{2^{k-1}}\bigg)
\]
where 
\[ \Psi(\ell_0) = \min\{\Omega\big(\frac{\sqrt{T(n)\ell_0}}{2^{k-1}}\big),\Omega\big(\frac{T(n-\ell_0)}{2^{k-1}}\big)\}. \]
\end{corollary}
\begin{proof}
There are three cases to consider.\\
\underline{Case 1:} Suppose $\ell_0 \leq T(n)/2$. Let $D':\{0,1,\ldots,T(n)\} \to \{1,-1\}$ be such that for any $z \in \{0,1\}^{T(n)}$, we have $D(|z|) = D'(|z|)$. By Theorem~\ref{theorem:main}, the complexity of $G^D_k$ is $\Omega(d/2^{k-1})$ where $d = \text{deg}_{1/3}(D')$. By Paturi's Theorem, $deg_{1/3}(D') \geq \sqrt{T(n)\ell_0(D')} = \sqrt{T(n)\ell_0}$ and so
\[ R^\epsilon_k(G^D_k) = \Omega\big(\frac{\sqrt{T(n)\ell_0}}{2^{k-1}}\big) \]
\underline{Case 2:} Suppose $T(n)/2 < \ell_0 \leq n/2$. We find a lower bound on the communication complexity of $G^{D_t}$ where $t = \ell_0 - T(n-\ell_0)/2$. Let $D'_t:\{0,1,\ldots,T(n-\ell_0)\} \to \{1,-1\}$ be such that $D'_t(|z|) = D_t(|z|)$. So by Theorem~\ref{theorem:main}, the complexity of $G^{D_t}_k$ is $\Omega(d/2^{k-1})$ where $d$ is the approximation degree of $D'_t$. We know 
\begin{eqnarray*}
D'_t(T(n-\ell_0)/2) &=& D_t(T(n-\ell_0)/2) \\
 &= &D(T(n-\ell_0)/2 + \ell_0 - T(n-\ell_0)/2)\\
& = & D(\ell_0)\\ 
&\neq& D(\ell_0 - 1)\\
& =& D'_t(T(n-\ell_0)/2 - 1).
\end{eqnarray*}
Thus by Paturi's Theorem, $\text{deg}_{1/3}(D'_t) \geq \sqrt{T(n-\ell_0)^2/2}$. This implies
\[ R^\epsilon_k(G^D_k) = \Omega\big(\frac{T(n-\ell_0)}{2^{k-1}}\big). \]
\underline{Case 3:} Suppose $\ell_0 = 0$ and $\ell_1 \neq 0$. The argument is similar to the one for Case  2. Consider $D_t$ where $t = n - \ell_1 - T(\ell_1)/2$. Let $D'_t:\{0,1,\ldots,T(\ell_1)\} \to \{1,-1\}$ be such that $D'_t(|z|) = D_t(|z|)$. As in case 2, one sees that $D'_t(T(\ell_1)/2) \neq D'_t(T(\ell_1)/2 + 1)$, so $deg_{1/3}(D'_t) \geq \sqrt{T(\ell_1)^2/2}$. Therefore,
\[ R^\epsilon_k(G^D_k) = \Omega\big(\frac{T(\ell_1)}{2^{k-1}}\big). \]
Combining these three cases, we get the desired result.
\end{proof}




\bibliographystyle{abbrv} 

\bibliography{Disjointness}

\end{document}